\documentclass[%
 reprint,
 amsmath,amssymb,
 aps,
pra,
]{revtex4-1}

\usepackage{graphicx}
\usepackage{dcolumn}
\usepackage{bm}
\usepackage{epstopdf}

\begin{document}

\preprint{APS/123-QED}

\title{Measurement of Ion Motional Heating Rates over a Range of Trap Frequencies and Temperatures}

\author{C. D. Bruzewicz}
\email{colin.bruzewicz@ll.mit.edu}
\author{J. M. Sage}
 \email{jsage@ll.mit.edu}
\author{J. Chiaverini}
 \email{john.chiaverini@ll.mit.edu}
\affiliation{Lincoln Laboratory, Massachusetts Institiute of Technology, Lexington, Massachusetts 02420, USA}

\date{\today}

\begin{abstract}

We present measurements of the motional heating rate of a trapped ion at different trap frequencies and temperatures between $\sim$0.6 and 1.5 MHz and $\sim$4 and 295 K. Additionally, we examine the possible effect of adsorbed surface contaminants with boiling points below $\sim$105$^{\circ}$C by measuring the ion heating rate before and after locally baking our ion trap chip under ultrahigh vacuum conditions. We compare the heating rates presented here to those calculated from available electric-field noise models. We can tightly constrain a subset of these models based on their expected frequency and temperature scaling interdependence. Discrepancies between the measured results and predicted values point to the need for refinement of theoretical noise models in order to more fully understand the mechanisms behind motional trapped-ion heating.

\end{abstract}

\pacs{Valid PACS appear here}
\maketitle


\section{Introduction}
Motional heating presents a major obstacle to high-fidelity two-qubit gate operations in large-scale trapped-ion quantum computation \cite{turchette2000heating,gaebler2012randomized,tan2013demonstration}. As ion traps have been made smaller in pursuit of a scalable quantum information processing architecture, the measured heating rates due to electric-field noise near the trap frequency have been found to increase dramatically with decreasing ion-to-trap distance~\cite{PhysRevLett.97.103007}. Techniques have, however, been demonstrated that greatly lower the measured heating rate. Specifically, cooling the trap electrodes to cryogenic temperatures has shown improvements of two orders of magnitude over room temperature heating rates \cite{PhysRevLett.97.103007,PhysRevLett.100.013001,chiaverini2014insensitivity}. \textit{In situ} ion bombardment of the trap electrode surface has also yielded a comparable 100-fold reduction in electric-field noise \cite{hite2012100,daniilidis2014surface,mckay2014ion}. Pulsed laser cleaning of the trap surface has given a smaller but significant decrease in the measured heating rate \cite{allcock2011reduction}. These results suggest that the heating process may be driven by thermally-activated surface effects from the nearby trap electrodes.  In order to more fully understand the underlying mechanisms, recent articles have called for further measurements of the trap-frequency and temperature dependence of this so-called anomalous motional heating \cite{hite2013surface,brownnutt2014ion}.

We present here a detailed study of the motional heating rates of a $^{88}$Sr$^{+}$ ion trapped in a surface-electrode ion trap over a range of trap temperatures and frequencies. We also show motional heating rates measured before and after locally heating the ion trap chip above $105^{\circ}\mathrm{C}$ under ultrahigh vacuum (UHV) conditions to examine the possible effect of residual contaminants on the trap surface in our unbaked vacuum apparatus. We compare the measured heating rates and their functional dependences to those predicted by available theoretical electric-field noise models.

\section{Experiment}
The details of the experimental apparatus have been described previously \cite{sage2012loading,chiaverini2014insensitivity}. We load our surface-electrode ion trap from a remotely-located magneto-optical trap (MOT) of neutral $^{88}$Sr. Pre-cooled atoms are pushed from the MOT by a resonant laser beam and are subsequently photo-ionized. A resultant $^{88}$Sr$^{+}$ ion is then Doppler cooled and confined at the ion trap location. The ion is trapped 50 $\mu$m from the segmented, linear, surface-electrode Paul trap using a combination of rf and DC electric fields. By adjusting the DC electrode voltages, the axial trap frequency of interest here can be varied from $\sim$0.6-1.5 MHz with radial frequencies in the range of 4-5 MHz.  The ion trap was fabricated by sputtering a 2 $\mu$m thick layer of niobium onto a sapphire substrate and was patterned using standard photolithographic techniques.

To reach cryogenic temperatures, the ion trap is connected by a weak thermal link to the cold stage of a vibration-isolated cryocooler capable of reaching a base temperature of approximately 3.5 K. Besides cooling the ion trap, the cryocooler also quickly provides UHV conditions without traditional high-temperature bakeout procedures. A resistive heater attached to the ion trap assembly permits local heating of the trap with minimal heating of the cryocooler cold stage. Hence, the temperature of the trap can be continuously varied from below 5~K to above 380~K while maintaining low background pressure $P\!<\!10^{-9}$ Torr. We have verified the accuracy of our silicon diode temperature sensor at low temperatures by observing the superconducting transition of the niobium trap near $T_{C}=9.16(4)$ K using an \textit{in situ} four-point resistance measurement. This result is in agreement with our \textit{ex situ} measurement of $T_{C}=\!9.2(1)$ K.

To measure the motional heating rate, we first prepare the $^{88}$Sr$^{+}$ ion in the ground state of its axial motional mode using Doppler cooling on the $S_{1/2}\!\to\!P_{1/2}$ optical transition at 422 nm followed by resolved sideband cooling using the $S_{1/2}\!\to\!D_{5/2}$ transition at 674~nm, as described in \cite{chiaverini2014insensitivity}. The ion then heats due to ambient electric-field noise for a variable delay time, and we measure its average vibrational occupation $\bar{n}$ using the sideband ratio technique \cite{PhysRevLett.75.4011} on the $S_{1/2}\!\to\!D_{5/2}$ transition, repeating each measurement 200 times for each probe frequency and delay time. The vibrational occupation as a function of delay time is then fit to a line to extract the heating rate $\dot{\bar{n}}$.

\section{Results}
\begin{figure}
\begin{center}
\includegraphics[width=\columnwidth]{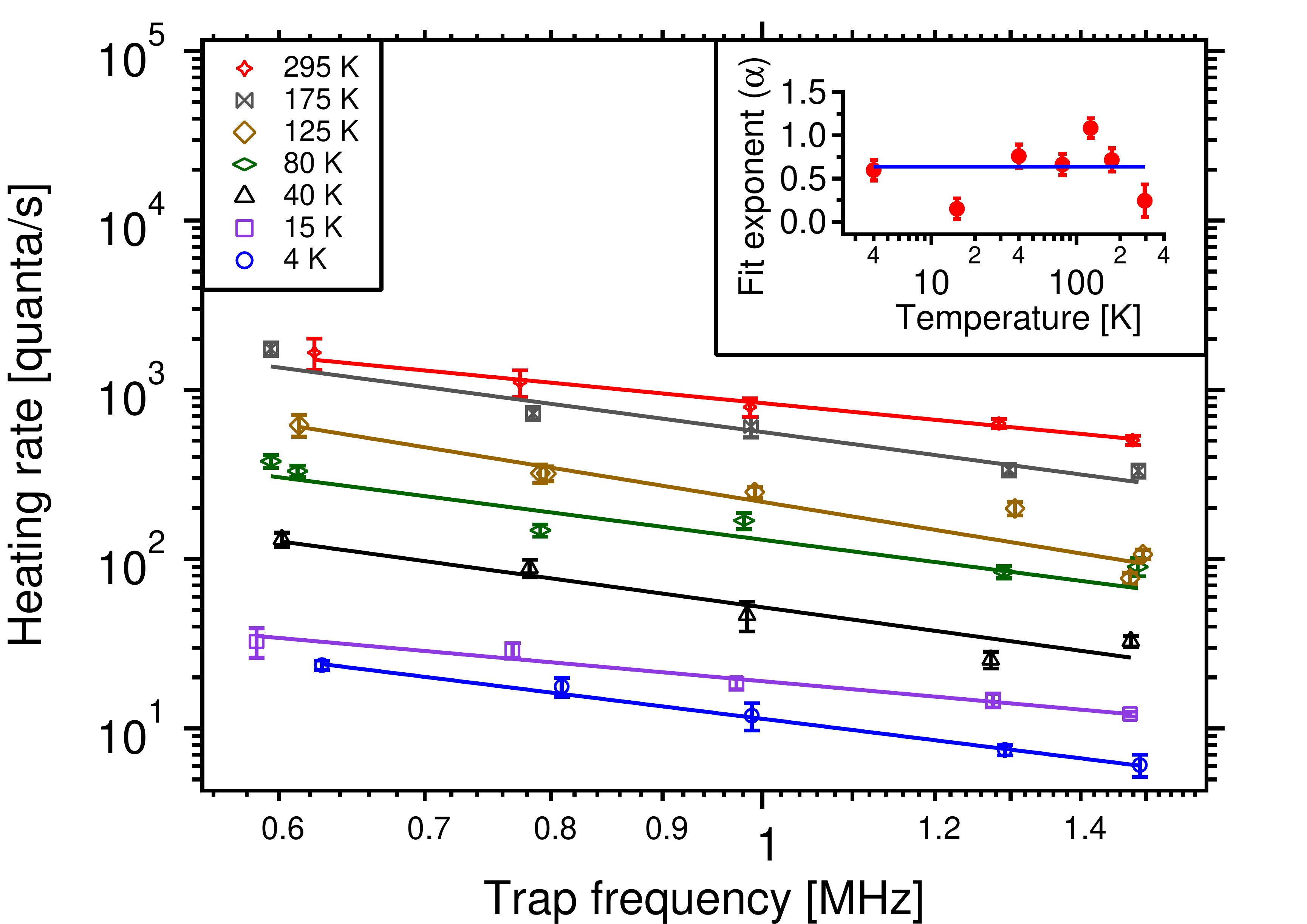} \\
\caption{Measured axial motional heating rate as a function of trap frequency for different trap temperatures. Error bars reflect statistical error propagated through the sideband ratio fitting procedure. Solid lines are fits to power law frequency scalings $\dot{\bar{n}}(f)\!\propto\!f^{-\alpha-1}$, which correspond to electric-field noise spectral densities $S_{E}(f)\propto\!f^{-\alpha}$. Inset shows the extracted fit exponents $\alpha$ with error bars from uncertainties of fits.  The weighted average value from this data $\alpha\!=\!0.6(1)$ is shown as a horizontal line.}
\label{fig:freq}
\end{center}
\end{figure}

\begin{figure}
\begin{center}
\includegraphics[width=\columnwidth]{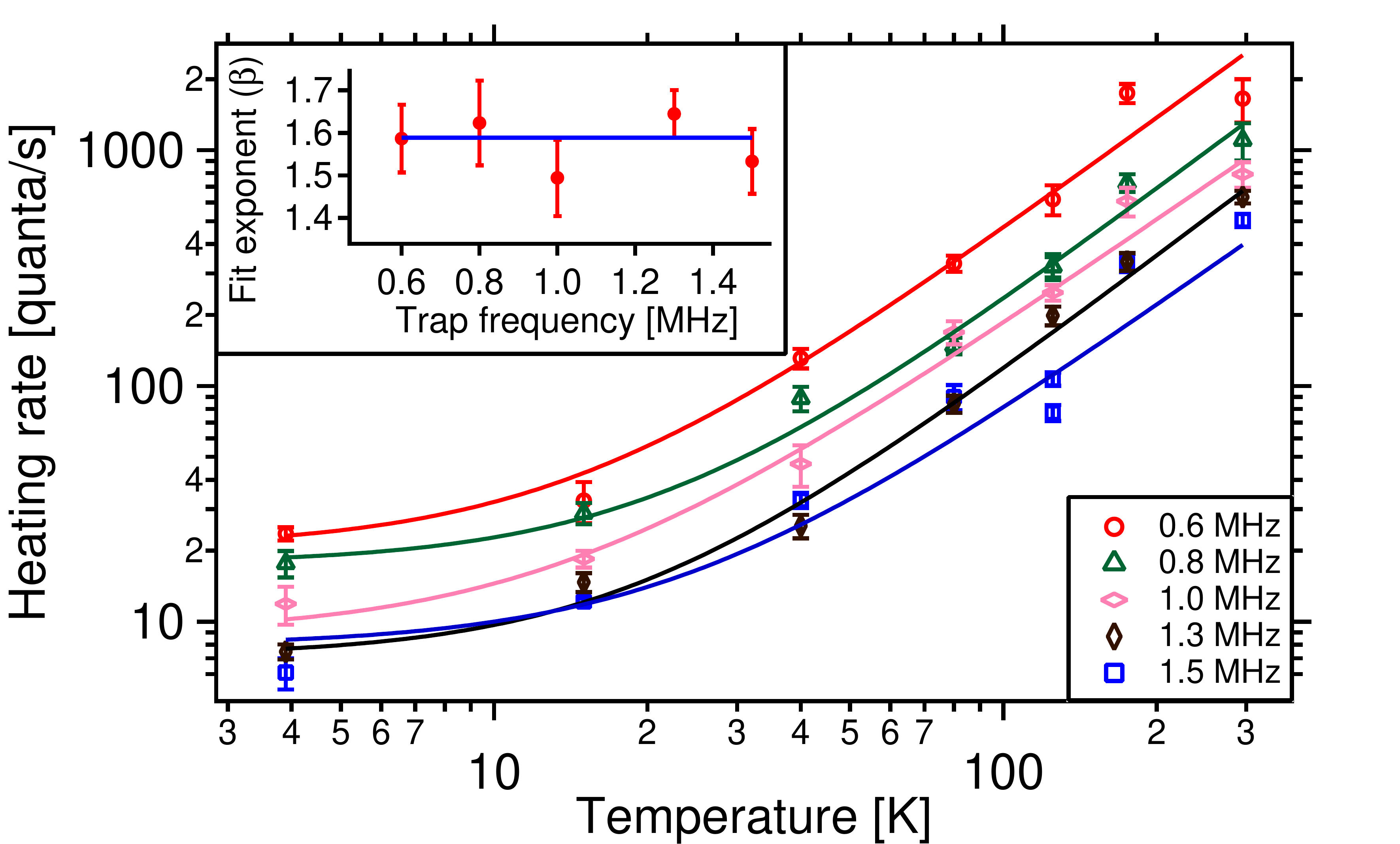}
\caption{Measured axial motional heating rate as a function of trap temperature for different trap frequencies. Solid lines are fits for power law temperature scaling with zero-temperature offset, $\dot{\bar{n}}(T)\!=\!\dot{\bar{n}}_{0}(1+(\frac{T}{T_{0}})^{\beta})$. Inset shows the extracted fit exponents $\beta$ with weighted average value of 1.59(3) shown as a horizontal line.}
\label{fig:temp}
\end{center}
\end{figure}

In Figures \ref{fig:freq} and \ref{fig:temp}, we present the measured heating rates of the axial trapping mode over a range of trap frequencies and temperatures between $\sim$0.6 and 1.5 MHz and $\sim$4 and 295 K. The motional heating rate $\dot{\bar{n}}$ is related to the electric-field noise spectral density $S_{E}(f)$ at ion trap frequency $f$ by
\begin{equation}
S_{E}(f)=\frac{4mhf}{q^{2}}\dot{\bar{n}},
\end{equation}
where $m$ is the ion mass, $h$ is Planck's constant, and $q$ is the electron charge \cite{wineland1998experimental}. We model the frequency dependence of the electric-field noise as  a single power law, $S_{E}(f)\!\propto\!f^{-\alpha}$. Although the heating rates vary by approximately two orders of magnitude over the measured temperature range, the fit exponents are approximately temperature-independent with no obvious trend. Hence, we report the weighted average value of $\alpha\!=\!0.6(1)$. These results are similar to those seen elsewhere in gold surface-electrode traps, where $\alpha$ values $\sim$0.7-0.8 were measured at multiple temperatures in the 30-75 K range \cite{PhysRevLett.101.180602}.

Analogously, the electric-field noise, and therefore the heating rate, also varies with the temperature of the ion trap \cite{PhysRevLett.97.103007,PhysRevLett.100.013001, PhysRevLett.101.180602, chiaverini2014insensitivity}. Here we model the temperature dependence of the heating rate for a fixed trap frequency as a power law with a non-zero offset at $T\!=\!0$. Specifically,
\begin{equation}
\dot{\bar{n}}(T)=\dot{\bar{n}}_{0}\bigg{(}1+\bigg{(}\frac{T}{T_{0}}\bigg{)}^{\beta}\bigg{)},
\end{equation}
where $\dot{\bar{n}}_{0}$ is the temperature-independent heating rate, $T_{0}$ is the thermal activation temperature, and $\beta$ is the high-temperature power law exponent.  The extracted fit parameters of this model are summarized in Table \ref{tab:tempvals}. The high-temperature scaling, and hence the power law exponent $\beta$, is remarkably insensitive to trap frequency over the measured range with a weighted average value of 1.59(3).

\begin{table}
\begin{center}
\renewcommand{\arraystretch}{1.5}
\begin{tabular}{|c|c|c|c|}
\hline
Trap Frequency [MHz] & $ \dot{\bar{n}}_{0}$ [quanta/s]& $T_{0}$ [K] & $\beta$ \\ \hline
0.60& 21(2) & 14(2)&1.6(1) \\ \hline
0.80&  18(2)& 21(4) & 1.6(1)\\ \hline
1.0& 9(2)& 13(4)& 1.5(1)\\ \hline
1.3& 7(1)& 19(2)& 1.6(1)\\ \hline
1.5& 8(1)& 24(4)& 1.5(1)\\ \hline
\end{tabular}
\caption{Fit parameters of temperature model,  $\dot{\bar{n}}(T)\!=\!\dot{\bar{n}}_{0}(1+(\frac{T}{T_{0}})^{\beta})$, for different trap frequencies extracted from the data in Figure \ref{fig:temp}. Uncertainties from the fits are given in parentheses.}
\label{tab:tempvals}
\end{center}
\end{table}

\begin{figure}
\begin{center}
\includegraphics[width=\columnwidth]{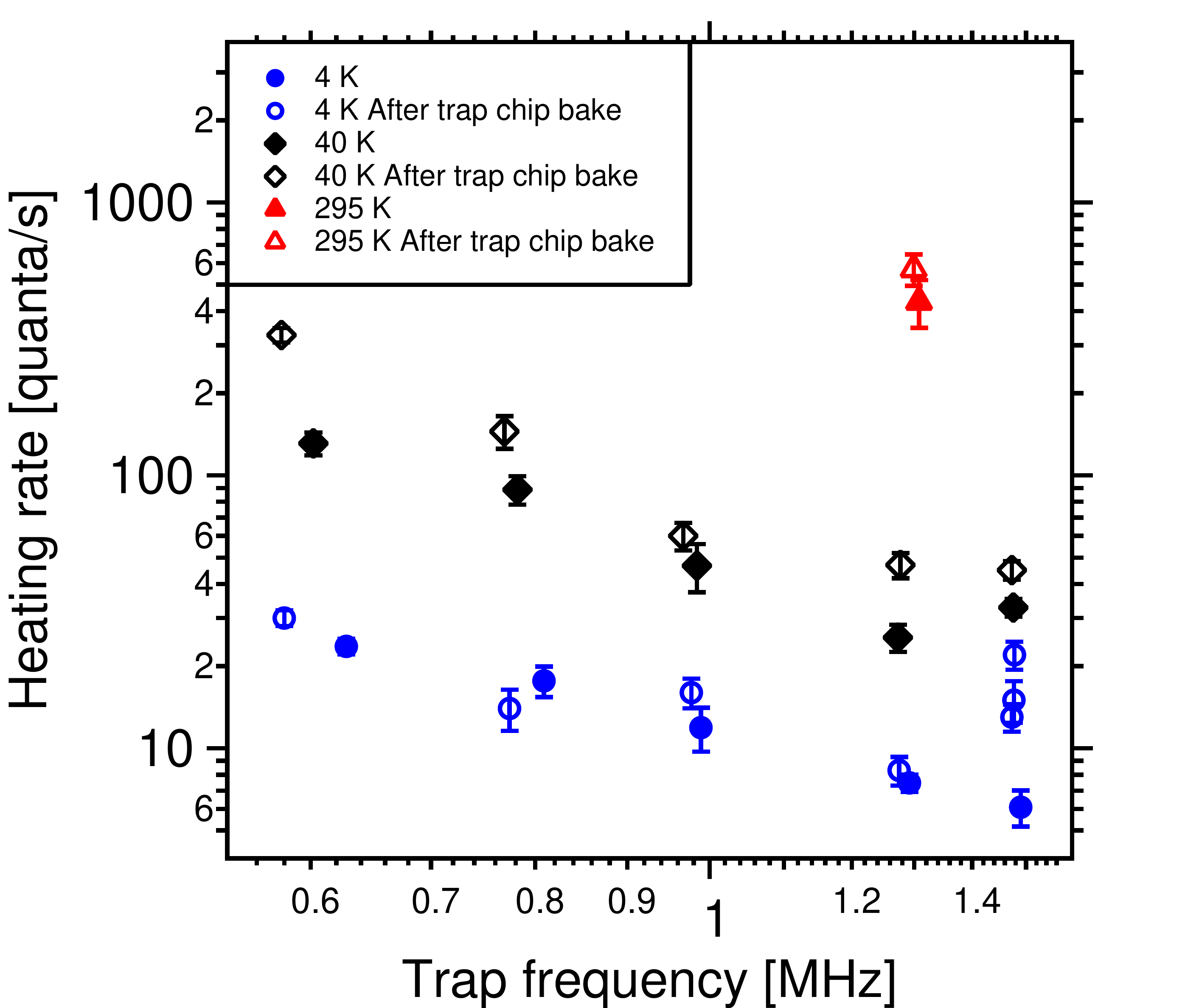}
\caption{Measured axial motional heating rates at different trap frequencies and temperatures before and after local baking of the ion trap chip at $107^{\circ}\mathrm{C}$ for $\sim$72 hours under UHV conditions. The frequency dependence of the heating rate and associated electric-field noise is largely unchanged following this procedure, suggesting that adsorbed water and laboratory solvents are not major causes of motional ion heating.}
\label{fig:minibake}
\end{center}
\end{figure}

Cryogenic vacuum systems obviate the need for traditional high-temperature bakeout procedures to reach UHV conditions. Although we maintain the trap chip at 295 K during the initial cooling of the cryocooler cold head to $\sim$4 K, it is possible that residual adsorbed contaminants, such as water or laboratory solvents, may remain on the trap surface. To investigate this process as a potential source of electric-field noise, we measured the motional heating rate at different trap frequencies and temperatures before and after locally baking the trap chip at $107^{\circ}\mathrm{C}$  for $\sim$72 hours while cryopumping and maintaining vacuum. This temperature is above the boiling points of water ($100^{\circ}\mathrm{C}$), methanol ($65^{\circ}\mathrm{C}$), ethanol ($78^{\circ}\mathrm{C}$), isopropyl alcohol ($83^{\circ}\mathrm{C}$), and acetone ($56^{\circ}\mathrm{C}$), permitting their removal from the trap surface under these conditions. Figure \ref{fig:minibake} shows that the frequency dependence of ion heating is largely unchanged following this trap chip bakeout procedure. Slight increases in the ion heating rate, especially at low temperature and high frequency, may be due to low-level contamination from increased outgassing and subsequent redeposition from trap stage materials with boiling points above the local bakeout temperature.

\section{Discussion}
Measurements of the heating rate at different trap frequencies and temperatures can be used to infer the functional dependence of $S_{E}(f,T)$, which can in principle be calculated for a given physical model of the electric-field noise. We examine available theoretical noise models and compare them to our measurements. These results are summarized in Table \ref{tab:models}.

\subsection{Johnson Noise}
In any conductor at finite temperature, thermal motion of electrons gives rise to Johnson noise. The corresponding electric-field noise spectral density is given by~\cite{brownnutt2014ion}
\begin{equation}
S^{\mathrm{(JN)}}_{E}(f,T)=\frac{4k_{B}TR(f,T)}{D^{2}},
\end{equation}
where $k_{B}$ is Boltzmann's constant, $R$ is the frequency and temperature-dependent resistance of the conductor, and $D$ is the characteristic length scale of the trap. For a planar surface-electrode trap, $D$ can be approximated as 2$d$, where $d$ is the ion-surface electrode spacing \cite{brownnutt2014ion}. The dominant source of resistance in our system, especially at low temperatures, is expected to be the Th\'{e}venin equivalent resistance $R_{eq}$ from the low-pass RC filters (cutoff frequency $\sim$16 kHz) on the electrode control lines. For the trap frequencies used here, $R_{eq}\!\propto\!f^{-2}$. In the 50-295 K temperature range, we calculate an approximate $R_{eq}\!\propto\!T^{0.3(1)}$ scaling due to the combined effects of the filters and the measured temperature-dependent resistance of the niobium electrodes. Hence, $S^{\mathrm{(JN)}}_{E}(f,T)\!\propto\!f^{-2}T^{1.3(1)}$, in contrast to our measured high-temperature $S_{E}(f,T)\!\propto\!f^{-0.6(1)}T^{1.59(3)}$ result. Finally, we can calculate the approximate magnitude of $S^{\mathrm{(JN)}}_{E}(f,T)$ for our experimental conditions. The predicted 4~K value of $S^{\mathrm{(JN)}}_{E}\!\approx\!10^{-17}\:\mathrm{V}^{2}/\mathrm{m}^{2}/\mathrm{Hz}$ greatly underestimates the measured values of $S_{E}\!\approx\!10^{-13}\: \mathrm{V}^{2}/\mathrm{m}^{2}/\mathrm{Hz}$ \cite{chiaverini2014insensitivity}. These deviations suggest that Johnson noise is not a major source of ion heating in our system.
 
\subsection{Adatom Diffusion}
Movement of adsorbates and their associated dipole moments across the electrode surface can change the electric-field at the ion location, giving rise to noise. This movement is generally described as a diffusion process with a diffusion constant $\mathcal{D}$ given by \cite{brownnutt2014ion}
\begin{equation}
\mathcal{D}=\mathcal{D}_{t}+\mathcal{D}_{0}e^{-V_{b}/k_{B}T},
\end{equation}
where $\mathcal{D}_{t}$ is a temperature-independent term that describes quantum tunneling and $\mathcal{D}_{0}$ is a thermally-activated component with activation energy $V_{b}$.  The corresponding electric-field noise spectrum is given by \cite{brownnutt2014ion}
\begin{equation}
S^{\mathrm{(AD)}}_{E}(f)=\frac{15\mu^{2}\bar{\sigma_{d}}\mathcal{D}}{16\pi\epsilon^{2}_{0}d^{6}(2\pi f)^{2}},
\end{equation}
where $\mu$ is the adatom dipole moment, $\bar{\sigma_{d}}$ is the average adatom density on the electrode surface, and $\epsilon_{0}$ is the permittivity of free space. This particular Arrhenius-type behavior predicts a temperature-independent baseline noise for $T\!\ll\!V_{b}/k_{B}$, followed by exponentially increasing noise until saturation for $T\!\gg\!V_{b}/k_{B}$. This is inconsistent with the measured power law with offset behavior seen in Figure \ref{fig:temp}, as the data does not fit well to an Arrhenius curve. The frequency dependence of $S^{(\mathrm{AD})}_{E}(f)$ for infinite planar geometries and needle traps have been calculated as $f^{-2}$ and $f^{-1.5}$, respectively \cite{brownnutt2014ion,wineland1998experimental}. Our segmented, surface-electrode trap geometry lies somewhere between these two limits, but the measured frequency dependence lies outside the predicted range of values. Further, the approximate magnitude of the electric-field noise has been calculated for reasonable experimental parameters for gold traps in Ref.~\cite{brownnutt2014ion}. For $\mu=5~$D, $\bar{\sigma_{d}}$=10$^{18}$ m$^{-2}$, $\mathcal{D}_{0}=10^{-7}$ m$^{2}$/s, $V_{b}/k_{B}=1750$ K, the predicted value for a planar trap at 295~K, where quantum tunneling effects should be negligible, is $S^{(\mathrm{AD})}_{E}\!\approx\!5\cdot10^{-16 }\:\mathrm{V}^{2}\!/\mathrm{m}^{2}\!/$Hz. This result is more than 2 orders of magnitude below our lowest measured values, which were taken at 4~K. All of these inconsistencies between the predicted and measured values suggest that adatom diffusion is not the primary source of motional ion heating.

\begin{table}
\begin{center}
\renewcommand{\arraystretch}{1.5}
\hyphenpenalty=10000
\begin{tabular}{|c|m{1.45cm}|m{2.05cm}|m{1.65cm}|}\hline
& Frequency Scaling& Temperature Scaling  & Approx. Magnitude $S_{E}(f,T)$ ($\mathrm{V}^{2}\!/\mathrm{m}^{2}\!/$Hz)\\ \hline
Johnson Noise & $f^{-2}$ & $T^{1.3(1)}$ &$1\cdot10^{-17}$ \\ \hline
Adatom Diffusion & $f^{-\frac{3}{2}}\!-\!f^{-2}$&$\mathcal{D}_{t}+\mathcal{D}_{0}e^{-T^{*}/T}$ & $5\cdot10^{-16}$ \\ \hline
Fluctuating Dipoles &$f^{0}-f^{-2}$ &$e^{-T^{*}/T}$ &  $2\cdot10^{-16}$  \\ \hline
Measured Results & $ f^{-0.6(1)}$ & $T^{1.59(3)}$& $1\cdot10^{-13}$ \\ \hline
\end{tabular}
\caption{Relevant functional scaling and approximate magnitudes for available electric-field noise models as well as the experimental results presented here for a 1.5 MHz trap frequency at 4 K. Approximate magnitudes for fluctuating dipole and adatom diffusion models come from parameter values for gold traps used in Ref. \cite{brownnutt2014ion}, scaled to our ion-electrode distance $d\!=\!50\:\mu$m. The adatom diffusion value is calculated for $T\!=\!295$ K for an activation energy $T^{*}\!=\!V_{b}/k_{B}\!=\!1750$ K. Quantum tunneling should be negligible at this temperature and can safely be ignored. The fluctuating dipole value is scaled to $T\!=\!4$~K for an activation energy $T^{*}\!=\!60$ K and fluctuation frequency $\Gamma_{0}\!=\!2\pi\!\cdot\!10$ MHz.}
\label{tab:models}
\end{center}
\end{table}

\subsection{Two-Level System Fluctuations}
Even in the absence of a physical mechanism, there exists a general formalism for describing the frequency spectrum of thermally-activated fluctuations in two-level systems \cite{RevModPhys.53.497}. With the inclusion of a distribution of activation energies, as opposed to a single energy scale, the temperature dependence departs from the exponential Arrhenius model. The exact form of the electric-field noise temperature and frequency scaling depends on the distribution of activation energies over the energy range of interest. For a given model distribution of activation energies, one can calculate the corresponding electric-field noise spectrum $S^{(\mathrm{TLS})}_{E}(f,T)$. The frequency scaling exponent $\alpha$ is related to the temperature scaling by \cite{RevModPhys.53.497}
\begin{equation}
\alpha(f,T)= 1- \frac{1}{\mathrm{ln}\hspace{0.25mm}(2\pi f\!\tau_{0})}\bigg(\frac{\partial\hspace{0.25mm}\mathrm{ln} \hspace{0.25mm}S^{(\mathrm{TLS})}_{E}}{\partial\hspace{0.25mm} \mathrm{ln}\hspace{0.25mm} T}-1\bigg),
\label{eq:alpha}
\end{equation}
where $\tau_{0}$ is a fluctuation ``attempt time" assumed to satisfy $f^{-1}\gg\tau_{0}$ and $\mathrm{ln}(2\pi f \tau_{0})\!\approx\!-10$. For a power law temperature dependence, $S^{(\mathrm{TLS})}_{E}(T)\!\propto\!T^{\beta}$, as seen in this work at high temperatures, Eq. \ref{eq:alpha} simplifies to 
\begin{equation}
\alpha(f,T)= 1- \frac{\beta-1}{\mathrm{ln}\hspace{0.25mm}(2\pi f\!\tau_{0})}.
\end{equation}
This result has the attractive feature that $\alpha$ is essentially frequency-independent, which is similar to the result in Figure \ref{fig:freq}, but for $\alpha<1$, we require $\beta<1$ (for $2\pi f\tau_{0}\!<\!1$). The high-temperature data in  Figure \ref{fig:temp}, however, yield a value of $\beta\!\approx\!1.6$ for all of the measured trap frequencies. Furthermore, if we constrain $S^{(\mathrm{TLS})}_{E}(T)$ only such that it increases monotonically with temperature, as seen here and in similar trapped-ion work \cite{PhysRevLett.97.103007,PhysRevLett.100.013001,chiaverini2014insensitivity}, we find
\begin{equation}
\alpha(f,T)>1+\frac{1}{\mathrm{ln}(2\pi f\!\tau_{0})}.
\end{equation}
Hence, for $\alpha=0.6(1)$ and $f=1$ MHz, we can constrain $\tau_{0}>5$ ns at the level of one standard deviation. This value is much larger than the typical attempt time suggested in \cite{RevModPhys.53.497}, which is on the order of $10^{-3}\!-\!10^{-5}$ ns.

The very large discrepancy in the constrained value of $\tau_{0}$ can be attributed to the significant deviation in the measured value of $\alpha$ below 1. If our experimental system were subject to technical noise at a level comparable to the electric-field noise at our highest trap frequencies, we would observe an artificial decrease in the extracted value of $\alpha$. Given that we see values of $\alpha$ significantly below~1 for several temperatures between 4-295 K, this proposed technical noise would have to change substantially and precisely in order to be comparable to the high trap frequency electric-field noise at each temperature, increasing by approximately 2 orders of magnitude over the measured temperature range. Such a pathological technical noise source seems very unlikely. Hence, we expect that the heating rates measured here are not primarily driven by thermally-activated two-level system fluctuations.

In the context of trapped-ion motional heating, an extension of the two-level fluctuation model that includes multiple levels has been proposed \cite{safavi2011microscopic}. In this model, thermally-activated, phonon-induced transitions between different bound adatom surface states lead to fluctuations in the magnitude of the dipole moment of the adatom-trap electrode system. At low temperatures, only the ground and first excited surface states can be populated. The electric-field noise spectral density for this two-level system is given by \cite{safavi2011microscopic}
\begin{equation}
S^{(\mathrm{FD})}_{E}(f,T)\simeq\frac{3\pi\sigma_{d}}{(4\pi\epsilon_{0})^{2}d^{4}}\frac{\Gamma_{0}(\Delta \mu)^{2}}{(2\pi f)^{2}+\Gamma_{0}^{2}}e^{-\hbar\nu_{10}/k_{B}T},
\label{eq:lowtemp}
\end{equation}
where $\sigma_{d}$ is the density of surface dipoles, $\Delta\mu$ is the difference in dipole moments of the surface states, $\Gamma_{0}$ is the decay rate from the excited to ground surface state at $T\!=\!0$, and $\hbar\nu_{10}$ is the energy splitting between the excited and ground surface states.  For typical values of $\Gamma_{0}\!=\!2\pi\cdot\!1$ THz \cite{brownnutt2014ion} and trap frequencies of $f\!\sim\!1$ MHz, the electric-field noise spectral density has essentially no frequency dependence.  At much higher frequencies, there would be a smooth transition from $f^{0}\!\to\!f^{-2}$ scaling, beginning when $2\pi f\!\sim\!\Gamma_{0}$. Although typically $2\pi f\!\ll\!\Gamma_{0}$, weakly-bound surface adsorbates, such as Ne, are predicted to yield fluctuation frequencies closer to experimentally attainable trap frequencies, on the order of a few MHz \cite{safavi2011microscopic}.

For very low temperatures, where $k_{B}T\!\ll\!\hbar\nu_{10}$, Eq.~\ref{eq:lowtemp} implies exponential suppression of dipole fluctuation and corresponding electric-field noise.  For temperatures where $k_{B}T\gtrsim\hbar\nu_{10}$, higher-lying surface levels beyond the first excited state can also be populated, changing both the temperature and frequency dependence of $S^{(\mathrm{FD})}_{E}(f,T)$. In this higher temperature regime, the crossover frequency from $f^{0}\!\to\!f^{-2}$ scaling becomes approximately equal to $\Gamma_{0}(1+(e^{\hbar\nu_{10}/{k_{B}T}}-1)^{-1})$ \cite{safavi2011microscopic}. Hence, this model predicts a trend of $\alpha$ values from $-2$ to 0 with increasing temperature if $k_{B}T_{\mathrm{min}}\lesssim\hbar\nu_{10}\lesssim \!k_{B}T_{\mathrm{max}}$, where $T_{\mathrm{min}} (T_{\mathrm{max}})$ is the lowest (highest) temperature used. The measured values in Figure \ref{fig:freq} are instead independent of temperature, clustered near $\alpha\!\approx\!0.6,$ suggesting that our measured temperatures are not in this crossover regime. 

In the high-temperature limit, where $k_{B}T\gg\hbar\nu_{10}$ for all relevant temperatures, we can extract an upper bound on $\nu_{10}$. Specifically, we require $\nu_{10}\!\ll\!k_{B}T_{\mathrm{min}}/\hbar$, which for $T_{\mathrm{min}}\!=\!4$ K, implies $\nu_{10}\!\ll\!2\pi\cdot\!80$ GHz. The vibrational frequency $\nu_{10}$ can be approximated by
\begin{equation}
\nu_{10}\approx\zeta\sqrt{\frac{U_{0}}{mz^{2}_{0}}},
\label{eq:nu}
\end{equation}
where $U_{0}$ is the surface state potential depth, $z_{0}$ is the equilibrium position in the potential, $m$ is the adsorbate mass, and $\zeta$ is a dimensionless factor of order unity~\cite{safavi2011microscopic,safavi2013influence}. To make the bound on $\nu_{10}$ as stringent as is reasonable, we consider the very weakly-bound Ne-Au system discussed in \cite{safavi2011microscopic}, which has a very low vibrational frequency due to its shallow surface binding potential. Using the available Ne-Au values for $U_{0}\!=\!10$~meV, $z_{0}\!=\!3$~\AA, and $\zeta\!=\!3.3$, we find values of $\nu_{10}\!\ll\!2\pi\cdot\!80$ GHz require adsorbate masses $m\!\gg\!500$ amu. Hence, in order to be in this high-temperature limit, the adsorbates must simultaneously bind as weakly as noble gases to the electrode surface while also being highly massive. In this limit, however, the crossover frequency from $f^{0}\!\to\!f^{-2}$ scaling is approximately equal to $\Gamma_{0}(1+k_{b}T/\hbar\nu_{10}),$ which is inconsistent with the observed frequency scalings in Figure~\ref{fig:freq}. 

An extension to this adatom dipole fluctuation model includes the effects of monolayer surface contamination on electric-field noise \cite{safavi2013influence}. Differences between the monolayer and the bulk metal can significantly alter the dipole fluctuation frequencies. In limiting cases of weakly-bound (physisorbed) He and strongly-bound (chemisorbed) N on a gold surface, $m\!\sim\!100$ amu adsorbates yield dipole fluctuation frequencies of $\Gamma_{(0,\mathrm{He})}\!\approx\!2\pi\!\cdot\!150$ MHz and $\Gamma_{(0,\mathrm{N})}\!\approx\!2\pi\!\cdot\!2.5$ GHz, which are still much larger than our trap frequencies. Hence, this monolayer model would predict frequency-independent heating rates for our trap parameters in both the physisorbed and chemisorbed limits.

Much more massive adatoms could give rise to a frequency-dependent electric-field noise spectrum. We can calculate the necessary adsorbate mass to reach dipole fluctuation frequencies on the scale of our trap frequencies, $f\!\sim\!1$ MHz, from \cite{safavi2011microscopic}
\begin{equation}
\Gamma_{0}\approx\frac{1}{4\pi}\frac{\nu^{4}_{10}m}{v^{3}\rho},
\label{eq:scale}
\end{equation}
where $v$ is the speed of sound in the electrode, and $\rho$ is the electrode metal density.  Using Eq.~\ref{eq:nu} and values of the density (8.57 g/cm$^{3}$) and speed of sound (5068 m/s)  in Nb, $\Gamma_{(0,\mathrm{He})}\!=\!2\pi\cdot\!1$ MHz requires $m\sim\!16000$ amu, and  $\Gamma_{(0,\mathrm{N})}=\!2\pi\cdot\!1$ MHz requires $m\sim\!270000$ amu. These masses are significantly higher than the values suggested in \cite{safavi2011microscopic,safavi2013influence,brownnutt2014ion}, which are on the order of 100 amu. Such large adsorbates are unlikely to be present in our vacuum system.

Although the present fluctuating dipole model does not reproduce our experimental results, it is possible that additional refinements of the theory may improve its predictive ability. Specifically, it would useful to examine the effect of combining the distributions of activation energies \cite{constantin2009saturation} and fluctuation frequencies of the two-state model with the multiple accessible surface states of the fluctuating dipole model. The resulting changes in the frequency and temperature dependences of the electric-field noise spectral density could possibly suggest a range of parameters that more successfully model our measurements. This extension of the model could be justified by the presence of multiple adsorbate species on the trap surface.

\section{Conclusion}
We have presented motional ion heating rates measured over a range of trap frequencies and temperatures. Johnson noise, adatom diffusion, thermally-activated two-level system, and  fluctuating dipole models have been examined, and all fail to reproduce our experimental results. We have also demonstrated that residual adsorbed water and laboratory solvents are not major causes of ion heating in our unbaked cryogenic apparatus. The lack of agreement between the measurements presented here and predictions from available noise models points to the need for additional theoretical and experimental work in order to understand and possibly overcome anomalous trapped-ion motional heating.

\begin{acknowledgments}
We thank Peter Murphy, Chris Thoummaraj, and Karen Magoon for assistance with ion trap chip packaging. We thank Robert McConnell for fruitful discussions and helpful editorial comments on the manuscript. This work was sponsored by the Assistant Secretary of Defense for Research and Engineering under Air Force Contract \#FA8721-05-C-0002. Opinions, interpretations, conclusions, and recommendations are those of the authors and are not necessarily endorsed by the United States Government.
\end{acknowledgments}

\bibliography{bibsputnb}
\end{document}